\documentclass{article}

\usepackage{arxiv}

\usepackage[utf8]{inputenc} 
\usepackage[T1]{fontenc}    
\usepackage{url}            
\usepackage{booktabs}       
\usepackage{amsfonts}       
\usepackage{nicefrac}       
\usepackage{microtype}      
\usepackage{lipsum}
\usepackage{graphicx}
\usepackage{tabularx}
\usepackage{booktabs}
\usepackage{tabularx}
\usepackage{multirow}
\usepackage{caption}
\usepackage{graphicx}

\usepackage{subcaption}

\usepackage[sort&compress, numbers]{natbib}
\usepackage{hyperref}
\usepackage{amsmath}

\title{Graph Neural Network Approach to Predict the Effects of Road Capacity Reduction Policies: A Case Study for Paris, France \thanks{Paper submitted for presentation at the 104\textsuperscript{th} Annual Meeting of the Transportation Research Board, Washington D.C., Jan. 2025}}

\author{
 Elena Natterer \thanks{Corresponding author} \\
  Chair of Traffic Engineering and Control \\
  Technical University of Munich, Germany \\
  \texttt{elena.natterer@tum.de} \\
   \And
 Roman Engelhardt \\
  Chair of Traffic Engineering and Control \\
  Technical University of Munich, Germany \\
  \texttt{roman.engelhardt@tum.de} \\
  \And
  Sebastian Hörl \\
  IRT System X, Paris, France\\
  \texttt{sebastian.horl@irt-systemx.fr} \\
    \And
 Klaus Bogenberger \\
  Chair of Traffic Engineering and Control \\
  Technical University of Munich, Germany \\
  \texttt{klaus.bogenberger@tum.de} \\
}

\begin{document}
\maketitle

\begin{abstract}
Rapid urbanization and growing urban populations worldwide present significant challenges for cities, including increased traffic congestion and air pollution. Effective strategies are needed to manage traffic volumes and reduce emissions. In practice, traditional traffic flow simulations are used to test those strategies. However, high computational intensity usually limits their applicability in investigating a magnitude of different scenarios to evaluate best policies. 
This paper introduces an innovative approach to assess the effects of traffic policies using Graph Neural Networks (GNN). By incorporating complex transport network structures directly into the neural network, this approach could enable rapid testing of various policies without the delays associated with traditional simulations.
We provide a proof of concept that GNNs can learn and predict changes in car volume resulting from capacity reduction policies. We train a GNN model based on a training set generated with a MATSim simulation for Paris, France. We analyze the model's performance across different road types and scenarios, finding that the GNN is generally able to learn the effects on edge-based traffic volume induced by policies. The model is especially successful in predicting changes on major streets. 
Nevertheless, the evaluation also showed that the current model has problems in predicting impacts of spatially small policies and changes in traffic volume in regions where no policy is applied due to spillovers and/or relocation of traffic.

\end{abstract}

\section{Introduction} \label{introduction}

The problem of urbanization presents significant challenges for cities worldwide, including increased traffic congestion and severe air pollution. These issues are exacerbated as urban populations grow, with 55\% of the world's population currently living in urban areas, a figure expected to rise to 68\% by 2050 \cite{un}. 
This rapid urbanization, combined with overall population growth, could add another 2.5 billion people to urban areas by mid-century \cite{wef_car_use}, further intensifying these problems. As cities continue to expand, they face the daunting task of managing increasing traffic volumes and the associated rise in emissions. Effective strategies are needed to reduce car use while ensuring flexibility for residents. Commonly employed policies include congestion charging, parking and traffic control measures, the establishment of limited traffic zones or rededicate space allocated for cars. 

Therefore, authorities must carefully decide which policy to introduce in which region of the city. In practice, traffic flow simulations are used to evaluate the impact of those policies. However, these simulations are typically unavailable for small to medium-sized cities as they require extensive data as input and expert knowledge for creation. Additionally, significant computational effort is required to run single simulations. This limitation hampers the ability to test a wide range of policies and determine the most effective ones, especially when considering district-specific implementations in large cities.

In this paper, we introduce a novel machine learning approach to assess the effects of traffic policies. By training a Graph Neural Network (GNN), complex network structures of the transportation system are directly incorporated into the neural network. The vision of this approach is to enable rapid testing of various policies without the lengthy delays associated with traditional simulations. By narrowing down the space of possibilities, a machine learning based model can offer a practical and efficient alternative for urban planners, facilitating more effective and timely decision-making in the management of urban transportation systems.

The goal of this paper is to provide a proof of concept if GNNs are capable of learning the impacts of policies, while a generalized model will be provided in future research.
We train and test the model on a transport simulation of Paris.
By conducting extensive simulations with capacity reduction policies in place, a database is created to learn resulting traffic flows on edge level in the network.
The results demonstrate that the trained GNN can accurately predict traffic flow under policies, proofing the applicability of GNNs for this task. Nevertheless, the GNN sometimes fails in predicting displacement traffic flow in regions outside applied policies.

This paper is organized as follows: The next section presents the literature review. Subsequently, we describe the developed method, i.e., the architecture of the GNN and used evaluation metrics. We then describe the case-study based on a MATSim simulation for Paris, France to test the method. Following this, we present the results. Finally, the conclusion highlights the insights gained, limitations of the proposed method, and future research directions.

\section{Literature Review}

To decrease car usage and thereby emissions and noise, numerous European cities implemented regulations either on neighborhood or even city scale~\cite{SadlerConsultantsEuropeGmbH.2024}. These regulations range from low emission zones, urban road tolls, parking restrictions or rededicating lanes to other modes of transport like public transport, bicycles, and pedestrian pathways. 
London and Stockholm are well known examples for urban toll systems in Europe, while the introduction of the tolling system for Manhattan, New York recently received general attention ~\cite{MTA.2024}.
In Paris, the city introduced numerous cycleways under the ``Vélo I and II'', consistently reducing at least one car lane to create ``Pistes cyclables''~\cite{Natterer.2024}.
Especially in the European Union, further regulation can be expected as new strict limits on air pollution have been set~\cite{EuropeanUnion.2023}. To cope with these challenges, methods need to be developed to evaluate the impacts of different regulations to sketch out ways to a sustainable urban mobility without restricting it.

A popular method to evaluate the impacts of regulations are agent-based simulations, like MATSim~\cite{horni_multi-agent_2016}, Polaris~\cite{Auld.2016}, SimMobility~\cite{Yang.2015} or mobiTopp~\cite{Mallig.2013}. The advantage of agent-based simulations is that the complex interaction between demand and supply in mobility systems can be represented with single agents being able to adopt their trip, i.e. by re-routing or mode choice adoptions if supply or demand configurations are changed to reach a new user equilibrium. Therefore, applications range from the evaluation of parking pricing schemes for Zurich, Switzerland~\cite{balac_modeling_2017}, low traffic zones for Paris, France~\cite{yin_evaluation_2024}, Superblocks (traffic management schemes in dense urban neighborhoods) in Vienna, Austria~\cite{straub_simulation_2023} and Barcelona, Spain ~\cite{barcelona_superblocks.2023}, or perimeter control measures for Christ Church, New Zealand. Also congestion pricing or tolling systems have been studied extensively, for example for New York, US~\cite{he_validated_2021} or Munich, Germany~\cite{Bracher.2017}.

The disadvantage of agent-based simulations is a high run-time of usually several hours, limiting the amount of scenarios that can be tested.
An alternative is to apply optimization based models to directly evaluate optimal policies (or design strategies) for a given network, usually referred to as Network Design Problem (NDP)~\cite{Boyce.1984, Magnanti.1984} traditionally evaluating the effect of network size on congestion and traveler choices with further extensions to public transport networks (e.g.~\cite{Scheele.1980, IbarraRojas.2015, Ng.2024}). A lot of research focus has been put in recent years in improving and extending these models.
This includes optimizing urban road space for multimodal networks (e.g.~\cite{Zheng.2013}), and developing optimal road pricing schemes (e.g.~\cite{Tirachini.2014,Loder.2022}). Nevertheless, as the these optimization-based models are notoriously hard to solve, strong modelling assumptions on user behavior simplifying the problem or rather small system sizes are applied. To combine a structured search for optimized policies with the advantages of agent-based simulations, Bayesian Optimization has been suggested to limit the search space for simulation runs (e.g.~\cite{Huo.2023, Dandl.2021}).

Shulajkovska et al. developed an open-data, open-source smart-city framework designed to enhance decision-making in European cities. Similar to this paper, their approach utilizes MATSim simulation outputs as a foundation for their machine learning algorithms. One of their key innovations is significantly speeding up policy testing for decision-makers, reducing the time required for a single policy verification from 3 hours to approximately 10 seconds. \cite{ml_for_pol_prediction}

The machine learning algorithms they tested include logistic regression, decision trees, and Bayesian ridge regression. However, these algorithms do not fully leverage the complex structures of urban networks. This study proposes using Graph Neural Networks (GNNs) to effectively manage intricate network structures for predicting traffic behavior at the edge level. 

Jiang et al.'s surveyed recent studies on traffic forecasting with GNNs, offering resources and identifying research challenges and opportunities in the field \cite{Weiwei2023}. Li et al. used Graph Convolutional Neural Networks to predict short-term citywide traffic demand using Graph Convolutional Networks (GCNNs). They developed a data-driven graph convolutional network (DDGCNN) that outperforms other predictors, particularly with weighted adjacency matrices, in capturing correlations between sub-regions \cite{Li.2020}. Lastly, Liu and colleagues introduce SimST, a spatio-temporal learning approach that models spatial correlations and achieves comparable performance to Spatio-Temporal Graph Neural Networks (STGNNs) while improving prediction throughput. This suggests that GNNs may not be the only effective option for spatial modeling in traffic forecasting \cite{Liu2023}.

While GNNs gained attention in recent years, they have been mainly applied for short-term traffic prediction.
This study evaluates the feasibility of learning long-term effects of traffic policies with GNNs potentially providing significant advantages for transportation planners in the evaluation of possible policies in a city.


\section{Method} 
\label{sec:method}

\begin{figure}
    \centering
    \includegraphics[width=1\linewidth]{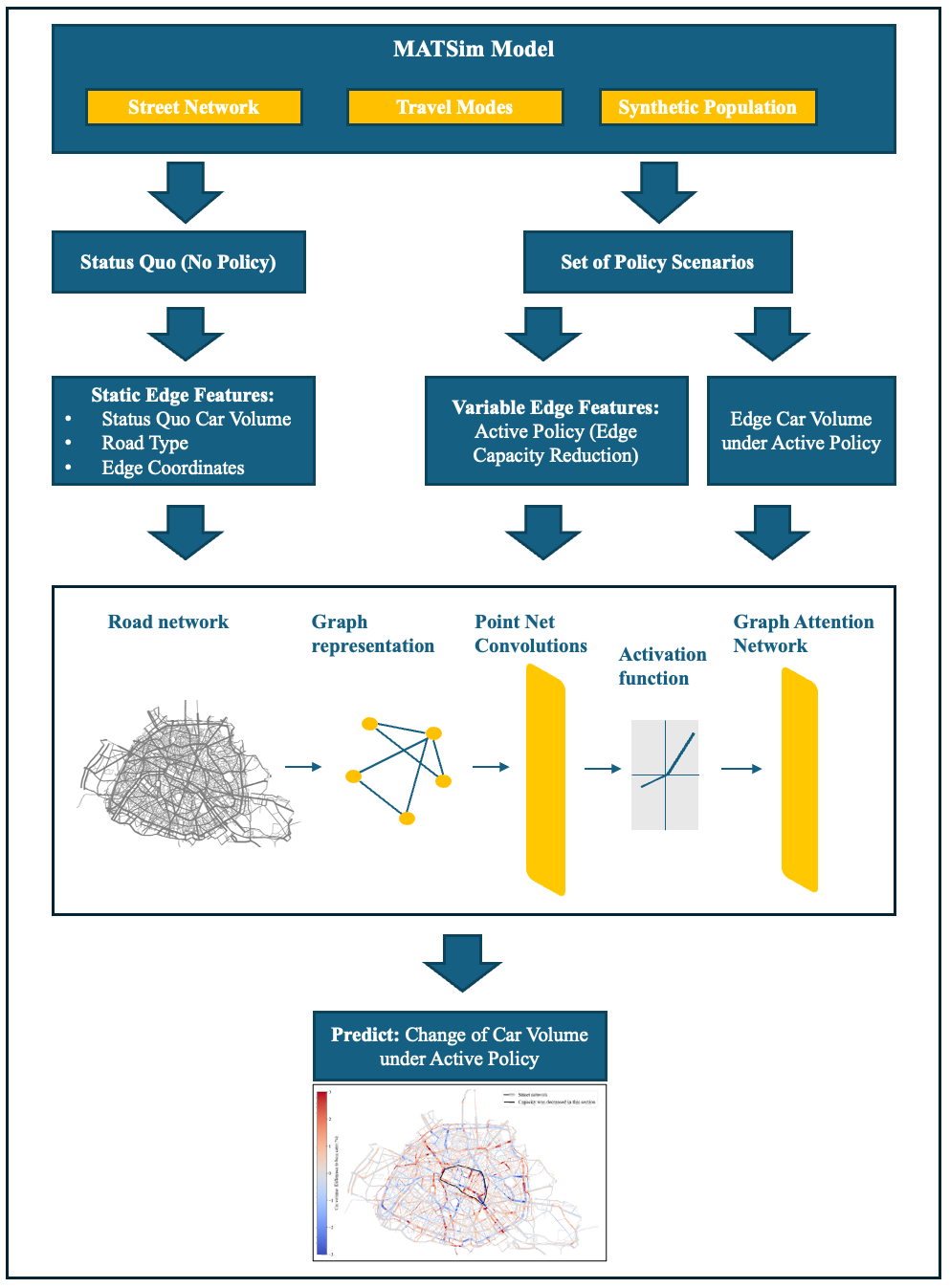}
    \caption{Method overview}
    \label{fig:flowchart}
\end{figure}

We propose the following method, as illustrated in Figure \ref{fig:flowchart}: An agent-based demand model (in this case, a MATSim simulation model) is used to train the Graph Neural Network (GNN). The road network defines the input graph to the GNN. 
Static (scenario independent) edge information include road type of each edge, coordinates of the midpoint to provide spatial information, and the car volume in the status quo, i.e. in the scenario without any policy applied.

To evaluate the general methodology, we focus on testing a policy involving the reduction of road capacity for motorized vehicle traffic. Capacity reduction can have different interpretations for real world implementation: For instance, that traffic-calming measures such as bumpers or street bays have been implemented. Link capacities can also be reduced by removing lanes, for example to rededicate space for additional bike lanes. Other measures could be the change of traffic light intervals, the introduction of crosswalks, the addition of parking spots, or the reduction of maximum speed. A reduction of link capacities can, hence, be seen as a generic representation of various traffic-related policies.

To create training data, different sets of policy scenarios (in our case, spatial variations of edge capacity reductions) are created, and the agent-based simulation is run. Under the active policy, agents in the simulation can take other routes and/or use different modes of transport for their trips to minimize their travel costs. Therefore, the simulation model can capture impacts of the policy, including but not limited to changes in car volumes per edge. 

Next to static edge features, the active policy per edge (in our case the reduction in edge capacity) is a variable (policy scenario dependent) input to the GNN. The goal of the GNN is to predict car volumes per edge. While the simulation output for different policy scenarios are used to train the GNN, the goal is to predict edge car volumes for unseen policy scenarios.




The approach generally allows learning the impacts of various types of policies by extending corresponding edge and/or node input attributes, as long as it is possible to implement it in the simulation model.


 A reduced capacity can mean, for instance, that traffic-calming measures such as bumpers or street bays have been implemented. Link capacities can also be reduced by removing lanes, for instance, for the installation of bike lanes. Other measures could be the change of traffic light intervals, the introduction of crosswalks, the addition of parking spots, or the reduction of maximum speed. A reduction of link capacities can, hence, be seen as a generic representation of various traffic-related policies.

\subsection{Network Representation}

Graph Neural Networks operate by taking multiple graphs as input; nodes represent entities and edges represent relationships between these entities. 
Through a series of message-passing layers, nodes aggregate information from their neighbors. This iterative process allows nodes to update their feature representations based on the structure of the graph and the features of their connected nodes, ultimately leading to a comprehensive embedding that captures both local and global graph information. These embeddings can then be used for various tasks such as node classification, link prediction, and graph classification.

In our use case, we aim to predict the change of car volume on all edges due to a given policy. Road networks of cities form a natural graph, where edges represent roads and nodes represent intersections. Since Graph Neural Network message-passing layers primarily focus on predicting node features rather than edge features, we propose using the line graph (or ``dual'') of the street graph. In this line graph, nodes correspond to road segments, and edges represent the connectivity between these road segments. The goal is to predict the change in car volume at the edge level after implementing specific policies. The steps are as follows:
\begin{enumerate}
    \item \textit{Policy Selection:} Decide on the policies to test and identify the districts where these policies will be applied.
    \item \textit{Scenario Generation:} Generate scenarios based on these policies to be simulated with a simulation tool of choice.
    \item \textit{Simulation Execution:} Run these scenarios using a simulation tool, such as MATSim.
    \item \textit{Graph Representation:} For each simulation output corresponding to a policy implementation, create a graph representation of the dual of the resulting street network. These graph representations together form the resulting dataset.
    \item \textit{GNN Training and Validation:} Split the dataset into training, testing, and evaluation sets. Feed the training and validation sets into the Graph Neural Network (GNN) for training and validation.
    \item \textit{GNN Testing:} Use the trained GNN to test other scenarios and predict the changes in car volumes on road segments due to the implemented policies.
\end{enumerate}

\subsection{Evaluation metrics}
\begin{table}[t!]
    \centering
    \caption{List of symbols used in this analysis.}
    \label{tab:symbols}
    \begin{tabularx}{\textwidth}{ c | X  }
    \toprule
    Symbol & Description \\
    \midrule
    $E$ & Set of edges in the network. Contains elements $e \in E$.\\
    $|E|$ & Number of edges $e \in E$.\\
    \midrule
    $b_e$ & Base volume car on edge $e \in E$ \\
    \midrule
    $y_e$ & Actual (simulated) change of car volume on edge $e \in E$ due to intervention \\
    $\hat{y}_e$ & Predicted change of car volume on edge $e \in E$ \\
    $\overline{y}$ & Mean actual change of car volume over all edges $e \in E$ \\
    \midrule
    $v_e$ & Actual car volume on edge $e \in E$ after intervention: $v_e = b_e + y_e$ for $e \in E$ \\
    $\hat{v}_e$ & Predicted car volume on edge $e \in E$ after intervention: $\hat{v}_e = b_e + \hat{y}_e$ for $e \in E$ \\
    $\overline{v}$ & Mean car volume on edge $e \in E$ after intervention \\
    
    \bottomrule
    \end{tabularx}
\end{table}

For training the Graph Neural Network and validating the results, we use two evaluation metrics: Mean Squared Error (MSE) and the coefficient of determination ($R^2$). We optimize the loss using MSE during training, while $R^2$ is used for additionally evaluating the model's performance. Both metrics are commonly used for regression tasks. 
Table \ref{tab:symbols} lists the variables used.

\subsubsection{Mean Squared Error}
The MSE for the change in car volume is defined in the following way:
  \begin{align}
    MSE(y, \hat{y}) = \frac{1}{|E|}  \sum_{e \in E} (y_e - \hat{y}_e)^2
\end{align}
$MSE(v, \hat{v})$ refers to the error predicting the overall car volume on edge level. 
Note that $MSE(y, \hat{y})$ is equivalent to $MSE(v, \hat{v})$: 
\begin{align}
\label{eqn:v_equation}
    MSE(v, \hat{v}) = \frac{1}{|E|} \sum_{e \in E} (v_e - \hat{v}_e)^2 = \sum_{e \in E} (b_e + y_e - b_e -\hat{y}_e  )^2 = \sum_{e \in E} (y_e - \hat{y}_e)^2
\end{align}
In other words, \textit{the error in predicting the change in car volume} due to an intervention is equivalent to \textit{the error in predicting the absolute car volume} after the intervention. \\

\subsubsection{Coefficient of determination ($R^2$)}
The coefficient of determination, $R^2$, is a common method to evaluate how well a regression model fits the actual data. For any edge $e \in E$, $y_e-\hat{y}_e$ is the residual of edge $e$. The sum of squares of the residuals is denoted as $SS_{res}$ (note that it is equivalent to MSE($y_e, \hat{y}$)):
\begin{align}
    SS_{res} =\frac{1}{|E|}   \sum_{e \in E} (y_e - \hat{y}_e)^2
\end{align}
We call the differences between the observed values and the mean of the observed values $SS_{tot}$:
\begin{align}
    SS_{tot} =\frac{1}{|E|}   \sum_{e \in E} (y_e - \overline{y})^2 
\end{align}
The ratio of these two quantities defines how well the predictions are in comparison to the mean. In its most general form, $R^2$ is defined as:
\begin{align}
\label{r_2_equation}
R^2 = 1 - \frac{SS_{res}}{SS_{tot}},
\end{align}
It is evident that if $\hat{y}_e = y_e$ for all $e \in E$, then $R^2 = 1$. Conversely, if $\hat{y}_e = \overline{y}$ for all $e \in E$, then $R^2 = 0$. Generally, if $R^2 > 0$, the model predicts values more accurately than the mean of $y$. Conversely, if $R^2 < 0$, the model's predictions are less accurate than the mean of $y$.

In general, a higher $R^2$ value indicates a better fit for the model. However, it is important to note that high $R^2$ values do not necessarily imply the model is the best. Therefore, we use $R^2$ in conjunction with the MSE to assess model performance comprehensively.

\subsubsection{Baseline}
\label{sec:baseline}
We have outlined the criteria for evaluating our model's predictions, focusing on MSE and $R^2$.
A baseline serves as a reference point for comparing model performance. It sets the minimum performance standard that a predictive model should meet to be considered effective. Baselines are typically derived from simple methods and vary based on the problem type. Since this is the first approach to this topic, no specific baseline has been established yet. Therefore, we use the mean baseline for both MSE and $R^2$. The baseline for $R^2$ is $0$: If $\hat{y}_e = \overline{y}$ for all $e \in E$, then $R^2 = 0$. For MSE, the baseline is:
\begin{align}
\label{baseline_eqn}
MSE(\overline{y}, \hat{y}) =\frac{1}{|E|}  \sum_{e \in E} (\overline{y} - \hat{y}_e)^2
\end{align}
Observe that this is equivalent to $MSE(\overline{v}, \hat{v})$ (see Equation \ref{eqn:v_equation}).

\subsubsection{Variance of Squared Difference}

The variance of the MSE measures the diversity of changes in car volume across all edges due to the introduced policy. Given $y_e$ as the actual values for all $e \in E$, and $\overline{y}$ as the mean of these. Then
\begin{itemize}
    \item $d_e = (y_e - \overline{y})^2$ is the squared difference for each edge $e \in E$.
    \item $\overline{d} =\frac{1}{|E|}  \sum_{e \in E} d_e$ is the mean of the squared differences.
\end{itemize}
Then the variance of the squared difference is:
\begin{align}
\text{VAR} = \frac{1}{|E|} \sum_{e \in E} (d_e - \overline{d})^2
\end{align}
A high variance indicates that the policy causes a wide range of changes in car volume on different edges. Conversely, a low variance suggests that the policy's effect on car volume is more uniform across the network. This implies that the policy has a similar impact on all edges. Since the policy is applied only to a few neighboring districts and streets, a uniform effect across the network is not expected. With a small impact of a policy, the variance (and the baseline MSE) is small. In such cases, machine learning models typically struggle to learn effects because clear signals are missing.



\section{Case-study} 
\label{sec:case_study}
We evaluated our model based on a MATSim simulation for Paris, France. In our study, the policy implemented reduces the capacity of higher-order roads by 50\%. Higher order roads are those roads which are classified as ``primary'', ``secondary'' or ``tertiary'' in OpenStreetMap, and the correspondings links, i.e. ``primary\_link''. The reduction of capacities along the network link can be interpreted in multiple ways in terms of real-world implementations.

In the following, first the simulation is described, followed by the generation of training data and finally the set-up of the GNN. 

\subsection{MATSim simulation}
The presented learning method can be used with various simulation tools. In the present case, the agent-based transport simulation framework MATSim \cite{horni_multi-agent_2016} has been chosen, as it is widely used for detailed transport modelling and standardized simulation data is available for a wide range of openly accessible used cases, such as for Paris \cite{horl_introducing_2021} or Berlin \cite{ziemke_matsim_2019}. Furthermore, MATSim is a highly flexible and modular framework that provides the functionality to simulate various elements of the transport system from specific services (on-demand mobility, micromobility, shared mobility, intermodality, ...) and policies (road pricing, access restrictions, ...) including their impacts and requirements related to traffic and travel behavior. MATSim performs simulations iteratively in two phases: The first phase performs a traffic simulation obtaining key information such as travel times per trip for all individual agent trips, while the second phase lets agents perform decisions related to their mobility plans. The present paper makes use of a specific configuration of MATSim in which discrete-choice models are used to simulate mode decisions of the agents \cite{horl_synthetic_2021}. The MATSim-based simulation originates from a large-scale agent-based transport simulation for the Île-de-France region around Paris, from which only the city perimeter has been cut. It is based on a standardized and replicable open-data process for the generation of a synthetic population for the region \cite{horl_synthetic_2021}. This synthetic population consisting of households, persons, and their daily activity patterns is transformed into a MATSim simulation including a detailed mode choice model based on French survey data \cite{horl_towards_2023}. Instantiations of the simulation have been used to study, for instance, the impact of autonomous taxis \cite{horl_dynamic_2019} and low traffic zones \cite{yin_evaluation_2024}. Since the purpose of the present paper is to present a first proof-of-concept of the learning process, down-sampled simulations of 0.1\% of the households with accordingly scaled network capacities are used, which take about 20 minutes to run on a standard machine. OpenStreetMap (OSM) provides as a foundation for the MATSim road network. Based on OSM data, the considered region comprises approximately 31,000 edges and 20,000 nodes.

\subsection{Generation of Training Data}

\begin{figure}[b!]
    \centering
    \includegraphics[width=\linewidth]{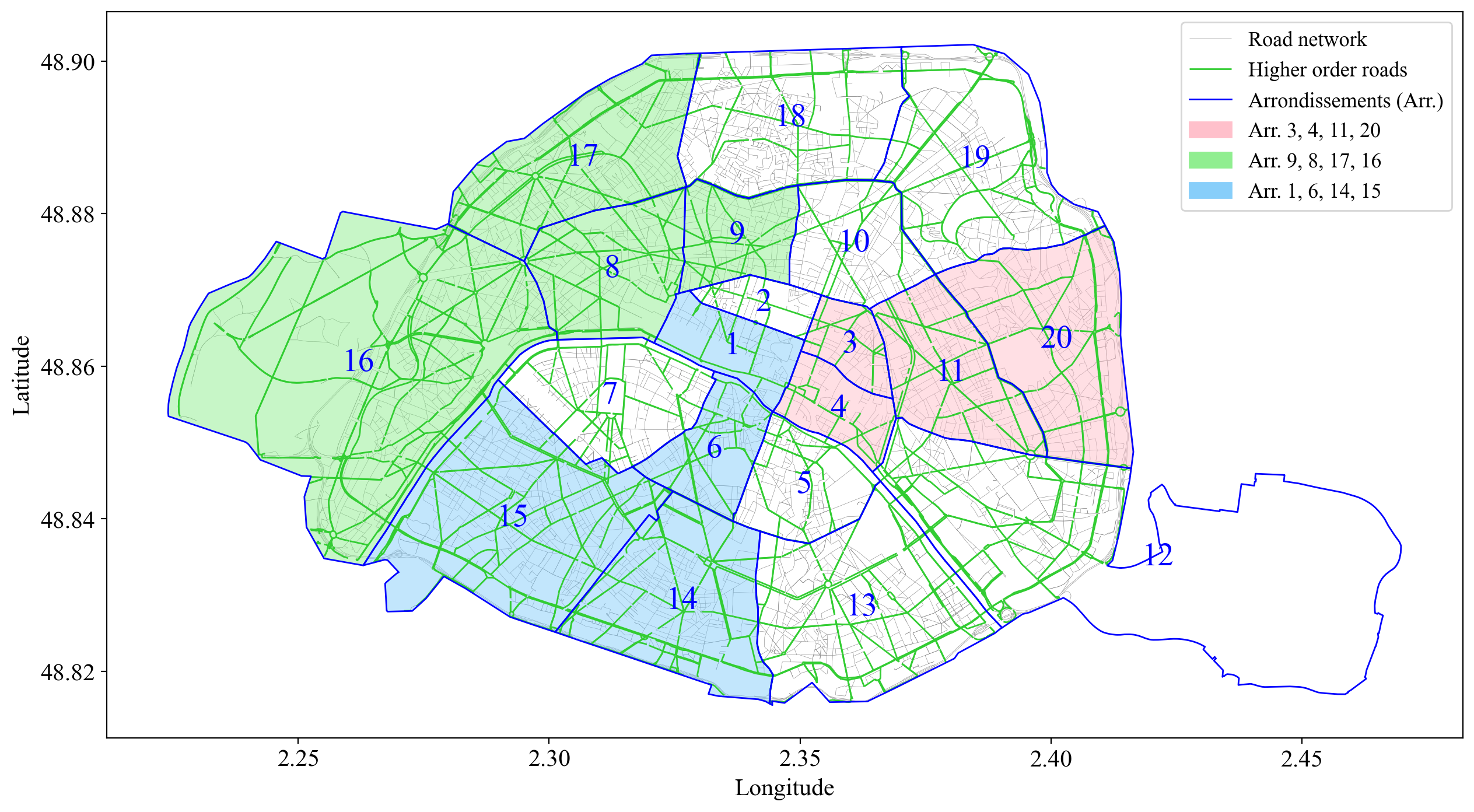}
    \caption{Simulation input: Policy is introduced in neighboring Arrondissements}
    \label{fig:sim-input}
\end{figure}

We created policy scenarios by reducing road capacities of street segements labeled as ``primary'', ``secondary'' or ``tertiary'' by 50\% in a given zone, which can be interpreted as removing lanes or reducing speed limits on main streets in the corresponding zone.
To provide a homogenous training set, various combinations of the 20 neighboring city districts (``Arrondissements'') in Paris are evaluated as zones for the policy scenarios:
To create connected zones, a graph is created connecting Arrondissements as nodes with their direct neighbours.
A depth-first search (DFS) is applied that ensures that each subset forms a path between every pair of nodes.
As it is not computational feasible to simulate all $368,048$ subsets, $4900$ randomly selected instances and all individual Arrondissements are simulated.
Note that there is a bias towards larger subsets, which may not be practical in implementation: Each Arrondissement appears in approximately half of the subsets, and on average, a zone is composed of 10 Arrondissements.
Figure \ref{fig:sim-input} illustrates three examples of simulation inputs, represented in red, green, and blue. 

Additionally to the policy scenarios, 50 simulation runs with different random seeds of the status quo scenario without policy are performed to evaluate a mean baseline car volume on each network edge.
Note that due to computational limits for each policy scenario, only one seed is computed, resulting in inherent stochastic variation of the training data because of using a 0.1\% travel demand subsample.

The different scenarios with their respective MATSim simulation outputs are randomly split into training, validation, and test sets with a ratio of 80\%, 15\%, and 5\%, respectively.

\subsection{Model set up and training process}
\label{sec:training_process}
The task of the GNN is to predict the change in car volume on an edge level due to a policy intervention. 
The features we use for each edge are as follows:
\begin{enumerate}
    \item Car volume in the base scenario
    \item Capacity in the base scenario
    \item Highway classification, encoded appropriately
    \item Geographic position
    \item Capacity reduction due to the policy
\end{enumerate}
We train our Graph Neural Network (GNN) using a carefully designed process to ensure stability and optimal performance. First, Standard Scaling normalization is applied to all feature variables, transforming the data so that each feature has a mean of 0 and a standard deviation of 1, which ensures consistent scaling across all features. The architecture of the network features a PointNetConvolution layer \cite{point_net_2017, point_net_2017_2} with two perceptrons: one for local features (consisting of a single linear layer of size 256) and one multilayer perceptron for global features (comprising four linear layers of sizes 256, 512, 256, and 512). This is followed by Graph Attention Network (GAT) layers \cite{gat_conv} with hidden sizes of 512, 512, 256, 128, and 64. The activation function we use is ReLu \cite{relu}. Overall, the resulting model has 833,411 parameters\footnote{The code for the Graph Neural Network can be found on \\https://github.com/enatterer/gnn\_predicting\_effects\_of\_traffic\_policies}. 

To ensure stability and avoid issues like vanishing or exploding gradients, we begin with weight initialization. We initialize the linear and GAT layers with Xavier Normal \cite{glorot2010understanding}, and the PointNetConv with Kaiming Normal \cite{he2015delving}.
The learning rate increases linearly during a warm-up phase of 20,000 steps to 0.001. After the warmup phase, we apply a cosine decay schedule \cite{sgd_warm_restarts} to smoothly decrease the learning rate, which ensures stable training. 

To manage memory constraints, we use a batch size of 8 and employ gradient accumulation, updating gradients every third step to simulate a larger batch size. Additionally, we use gradient clipping with a maximum norm of 1 to prevent exploding gradients.

We optimize the model using the AdamW optimizer \cite{weight_reg}, with a weight decay (set at 1e-3) for better regularization. 

\subsubsection{Training process}

The training runs for up to 2000 epochs, with early stopping if there is no improvement for 50 consecutive epochs, ensuring that the training concludes only when the model is fully optimized.
We evaluate the method outlined in section \ref{sec:method} using the case study described in section \ref{sec:case_study}.
The baseline for the model is 3.94 for the MSE, as stated in Equation \ref{baseline_eqn}, and $0$ for the coefficient of determination $R^2$ (see section \ref{sec:baseline}).

\begin{figure}
    \centering
    \includegraphics[width=1\linewidth]{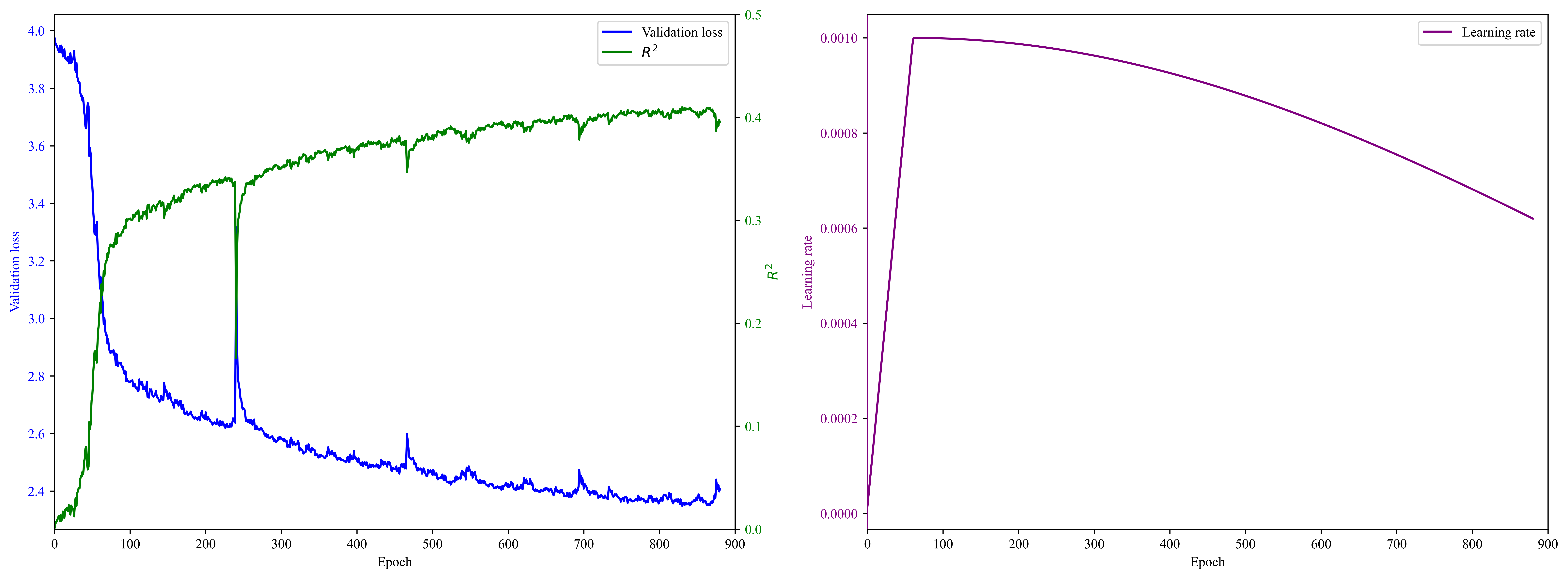}
    \caption{Visualization of the training process}
    \label{fig:training_process}
\end{figure}

The training process is depicted in Figure \ref{fig:training_process}. Initially, the validation loss is around $4$, similar to the baseline. It drops to $3$ after $60$ epochs and to $2.6$ after $220$ epochs. Training slows, reaching $2.5$ by $400$ epochs and stabilizing at $2.4$ by $880$ epochs. The $R^2$ score starts at $0$, increasing to $0.3$ after $90$ epochs, $0.35$ after $300$ epochs, and flattening at $0.4$ by $750$ epochs. The learning rate begins at $0$, reaching $0.001$ by epoch $60$ and then decreases due to the cosine decay scheduler, dropping to $0.0006803$ after $800$ epochs.

The training process converges after $880$ epochs with a final validation loss of $2.40$ and a final $R^2$ score of $0.41$. The training lasted for $26$ hours, on a single NVIDIA RTX A5000.

\section{Results} \label{results}

\begin{table}[htbp]
    \centering
    \caption{Result table, test set (245 observations)}
    \label{tab:result_table}
    \begin{tabularx}{\textwidth}{ p{6cm} | X | X | X | X | X }
    \toprule
    \textbf{Road subset} &  \textbf{Length (km)}
    &  \textbf{Variance} &
    \textbf{MSE: Baseline} & \textbf{MSE: Model} & \textbf{$R^2$} \\
    \toprule
    All roads & 2833 & 311.06 & 4.00 & 2.36 & 0.41 \\
    \midrule
    \midrule
    Roads of type primary & 356 & 726.39 & 10.15 & 5.27 & 0.48 \\
    \midrule
    Roads of type secondary & 299 & 106.29 & 4.23 & 2.97 & 0.30 \\
    \midrule
    Roads of type tertiary & 284 & 98.88 & 3.38 & 2.28 & 0.32 \\
    \midrule
    Roads of type primary, secondary or tertiary & 940 & 364.55 & 6.33 & 3.68 & 0.42 \\
    \midrule
    Roads of types other than primary, secondary, tertiary & 1894 & 260.41 & 2.29 & 1.39 & 0.39 \\
    \midrule
    \midrule
    Roads with capacity reduction & 329 & 130.12 & 5.31 & 3.73 & 0.30  \\
    \midrule
    Roads without capacity reduction & 2505 & 350.49 & 3.78 & 2.12 & 0.43 \\
    \midrule
    \midrule
    Roads of types primary, secondary, tertiary, and capacity reduction & 329 & 130.12 & 5.31 & 3.73 & 0.30 \\
    \midrule
    Roads of types primary, secondary, tertiary, and no capacity reduction & 611 & 486.56 & 6.85 & 3.65 & 0.47 \\
    \bottomrule
    \end{tabularx}
\end{table}

\subsection{General insights}
Applying the training process to our test set, we achieve an MSE of $2.35$ and an coefficient of determination $R^2$ of 0.41. Table \ref{tab:result_table} presents the evaluation metrics for the model's performance across various scenarios. The table includes six columns: The road subset, on which the model is evaluated, the length (in km) of roads where a capacity reduction has been implemented, the variance of that road subset, the baseline MSE, the MSE that our model achieves, and $R^2$. The values for columns 2 - 6 are each averaged over the 245 graph observations in the test set. 

The ``Variance'' column shows the variance of the data in each road set. The ``MSE: Baseline'' column presents the MSE for the baseline (see Equation \ref{baseline_eqn}), while the ``MSE: Model'' column presents the MSE for the trained model. The ``$R^2$'' column shows the coefficient of determination $R^2$, our metric used for evaluating how well a regression model fits the actual data. Note that for all scenarios, the baseline $R^2$ is $0$ (see section \ref{sec:baseline}).

For the entire road network (2,833 km), the variance is 311.06, with a baseline MSE of 4.00 and a model MSE of 2.36, resulting in an \(R^2\) of 0.41. In primary roads (356 km) with a high variance of 726.39, the baseline and model MSEs are 10.15 and 5.27, respectively, yielding an \(R^2\) of 0.48. Higher variance improves \(R^2\) by clarifying the impact of capacity changes. Secondary roads (299 km) show a variance of 106.29, with baseline and model MSEs of 4.23 and 2.97, and an \(R^2\) of 0.30. Tertiary roads (284 km) have a variance of 98.88, baseline MSE of 3.38, model MSE of 2.28, and an \(R^2\) of 0.32. Combined primary, secondary, and tertiary roads (940 km) have a variance of 364.55, baseline MSE of 6.33, model MSE of 3.68, and an \(R^2\) of 0.42. Roads of other types (1894 km) show a variance of 260.41, baseline MSE of 2.29, model MSE of 1.39, and an \(R^2\) of 0.39. For roads with capacity reduction (329 km, variance 130.12), baseline and model MSEs are 5.31 and 3.73, with an \(R^2\) of 0.30. Roads without capacity reduction (2505 km, variance 350.49) have baseline and model MSEs of 3.78 and 2.12, and an \(R^2\) of 0.43, showing better performance without capacity changes. In primary, secondary, and tertiary roads with capacity reduction (329 km, variance 130.11), baseline and model MSEs are 5.31 and 3.73, with an \(R^2\) of 0.30. Those without capacity reduction (611 km, variance 486.56) have baseline and model MSEs of 6.85 and 3.65, with an \(R^2\) of 0.47, highlighting improved model performance with higher variance and no capacity reduction.

In conclusion, the model performs best on primary roads, likely due to the higher traffic volume, which show higher impacts from applied policies and allows for more accurate predictions considering less stochastic variation due to downscaled 0.1\% simulations. It is also expected that the model achieves better performance on roads without capacity reduction, as only limited changes in traffic volume from the applied policies are observed for these sections. Generally, higher baseline MSE and variance correlate with increased prediction accuracy, indicating that the model's predictions surpass the baseline. This is anticipated, as greater variance suggests that the effects of policy changes are more distinctly captured in the simulation outputs.

\subsection{Insights for selected zones}

\begin{figure}[htbp]
    \centering
    \begin{subfigure}[b]{1\textwidth}
        \centering
        \includegraphics[width=1\textwidth]{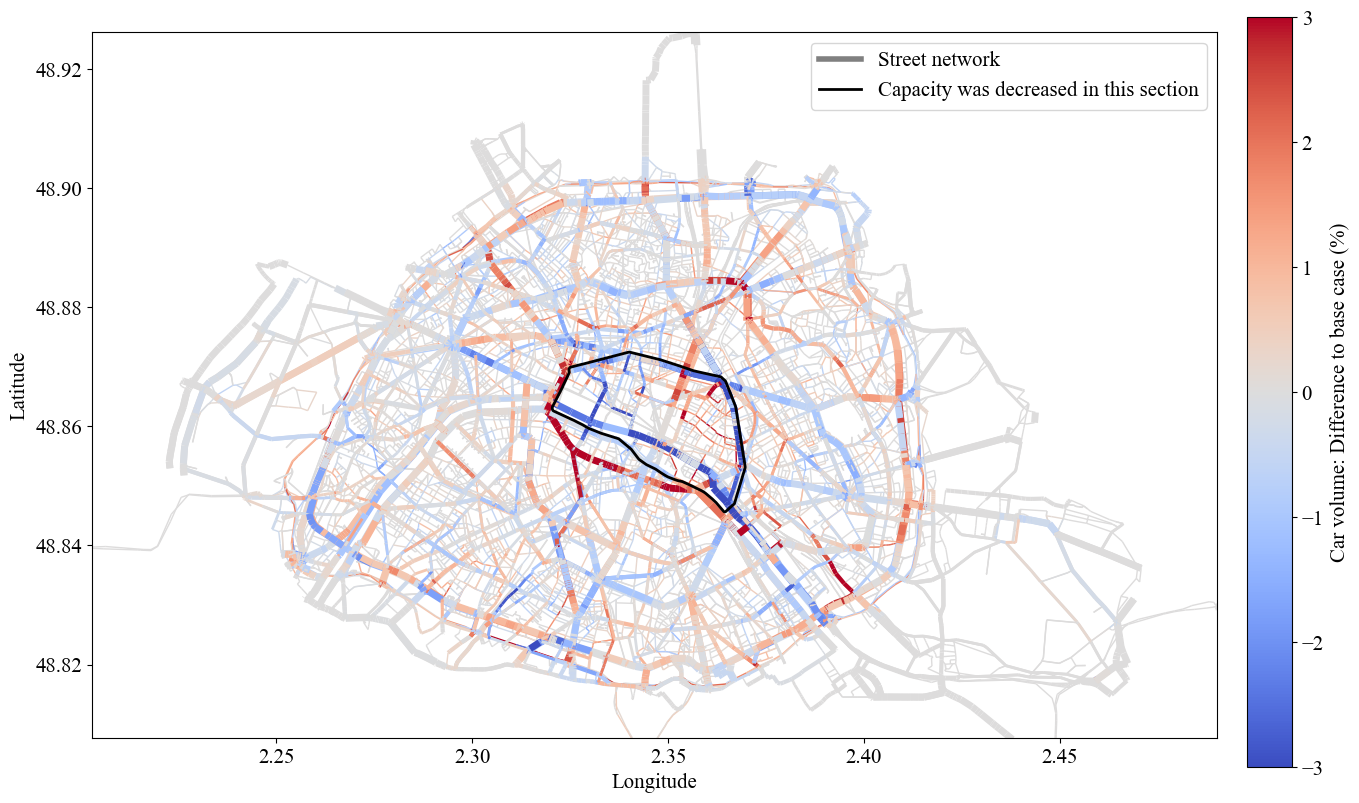}
        \caption{Zone 1: Actual change in car volume}
    \end{subfigure}
    \vfill
    \begin{subfigure}[b]{1\textwidth}
        \centering
        \includegraphics[width=1\textwidth]{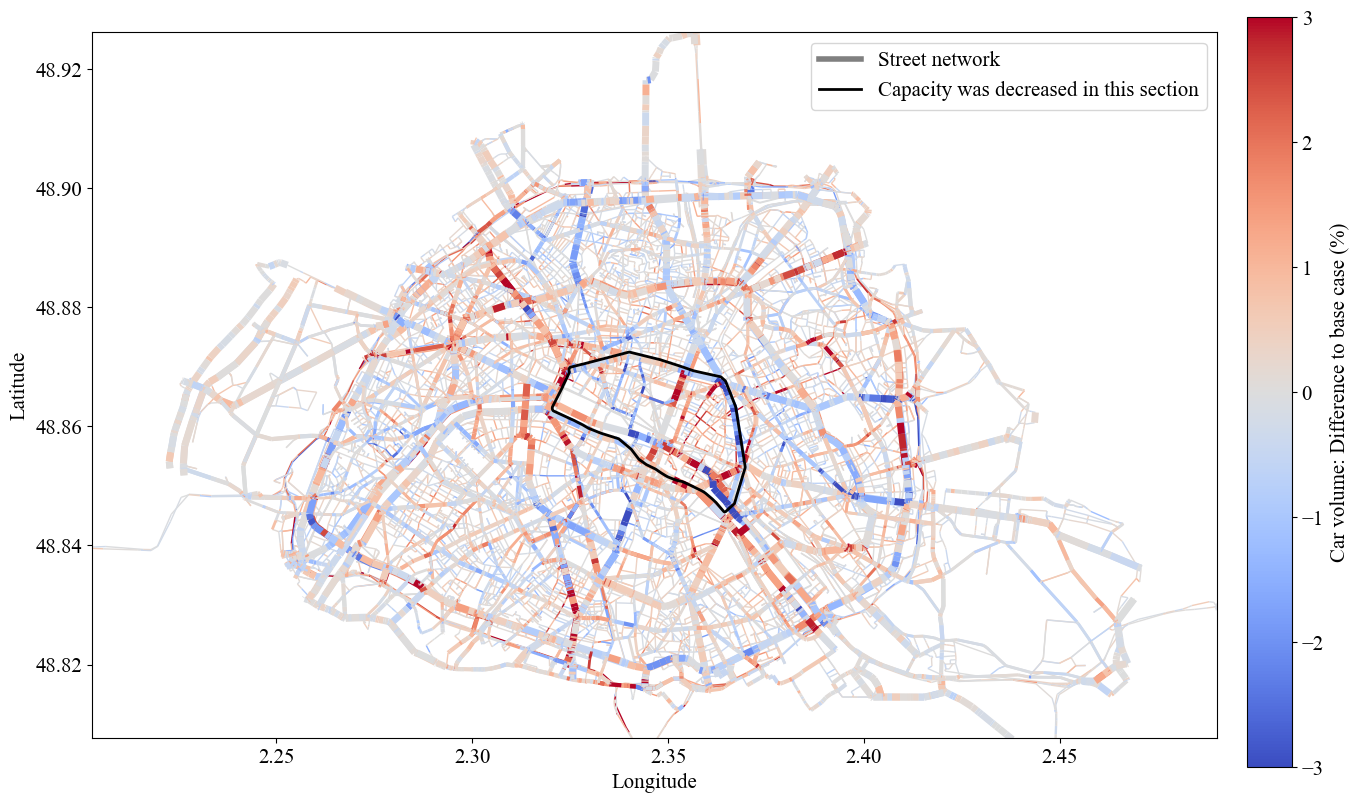}
        \caption{Zone 1: Predicted change in car volume}
    \end{subfigure}
    \caption{Comparison of actual and predicted change in car volume for Zone 1}
    \label{fig:zone_1_comparison}
\end{figure}

\begin{figure}[htbp]
    \centering
    \begin{subfigure}[b]{1\textwidth}
        \centering
        \includegraphics[width=1\textwidth]{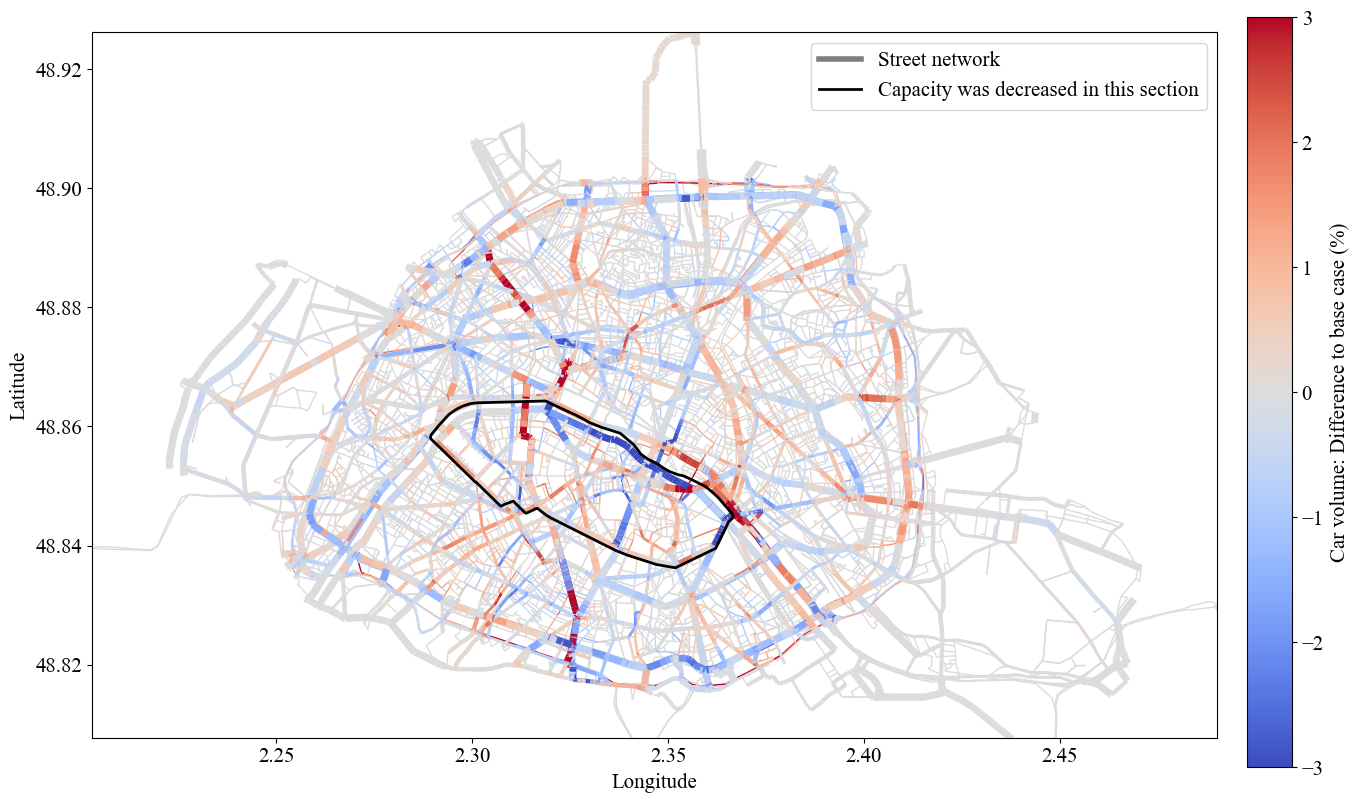}
        \caption{Zone 2: Actual change in car volume}
    \end{subfigure}
    \vfill
    \begin{subfigure}[b]{1\textwidth}
        \centering
        \includegraphics[width=1\textwidth]{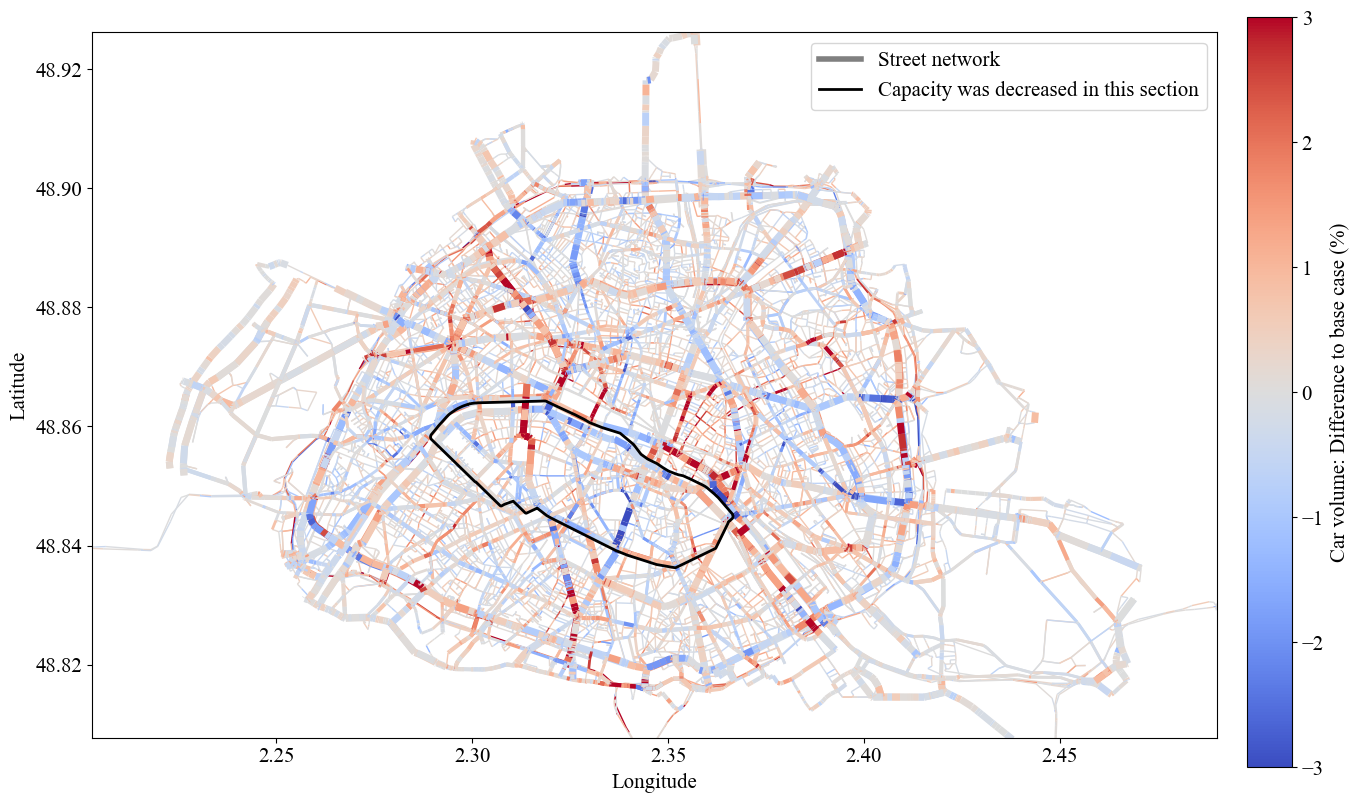}
        \caption{Zone 2: Predicted change in car volume}
    \end{subfigure}
    \caption{Comparison of actual and predicted change in car volume for Zone 2}
    \label{fig:zone_2_comparison}
\end{figure}

\begin{figure}[htbp]
    \centering
    \begin{subfigure}[b]{1\textwidth}
        \centering
        \includegraphics[width=1\textwidth]{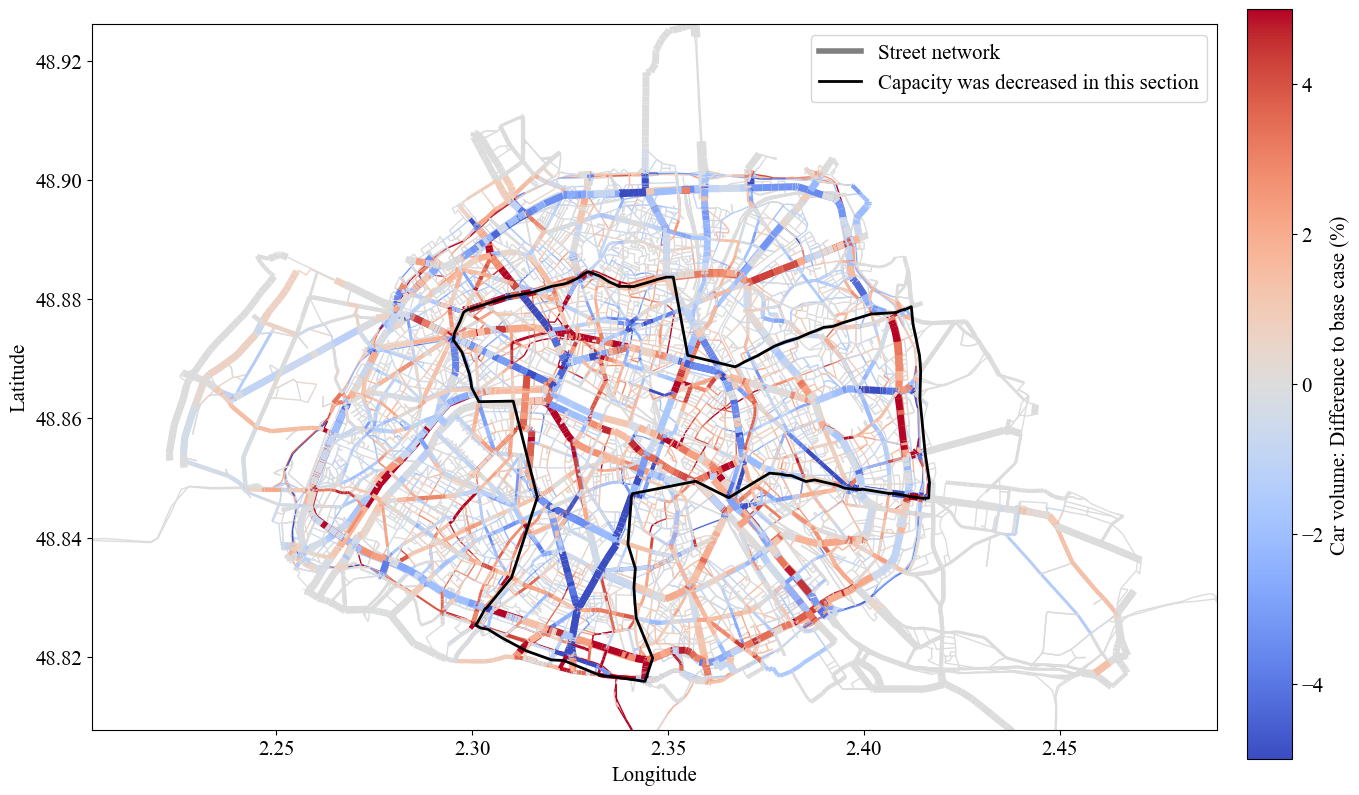}
        \caption{Zone 3: Actual change in car volume}
    \end{subfigure}
    \vfill
    \begin{subfigure}[b]{1\textwidth}
        \centering
        \includegraphics[width=1\textwidth]{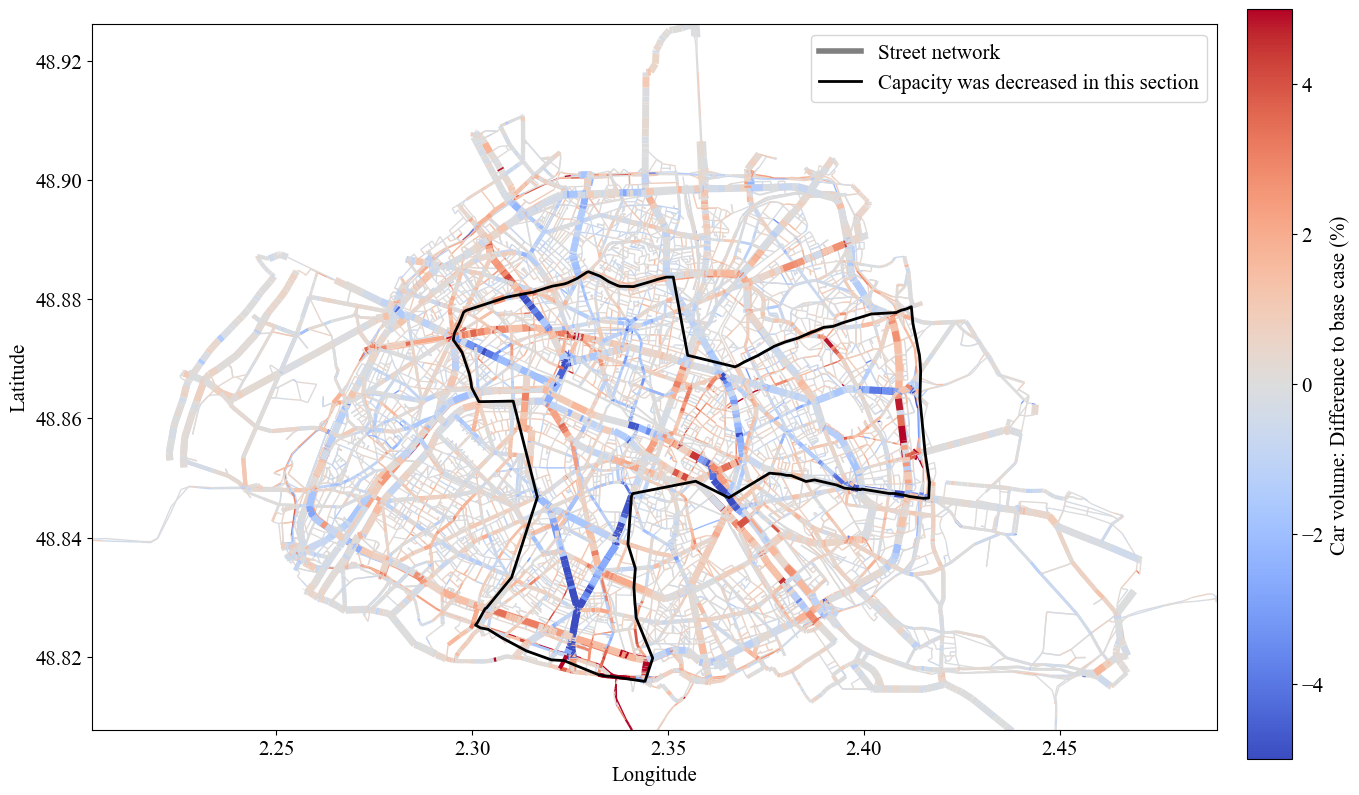}
        \caption{Zone 3: Predicted change in car volume}
    \end{subfigure}
    \caption{Comparison of actual and predicted change in car volume for Zone 3}
    \label{fig:zone_3_comparison}
\end{figure}

In this section, we demonstrate the results for three selected zones: Zone 1 encompasses Arrondissements 1 to 4. These central districts, situated north of the Seine, represent the centre of Paris. Zone 2 comprises Arrondissements 5 to 7. Located south of the Seine, these districts are also quite central and have undergone significant revitalization efforts in the last decade
Lastly, Zone 3 consists of Arrondissements 1, 2, 3, 4, 6, 8, 9, 11, 14 and 20. This combination spans across the central, eastern, and southern parts of the city.
In recent years, Paris has experienced substantial changes of its road network supply, with the goal of drastically reducing emissions \cite{TrafficReductionParis, Natterer.2024}. In reality, Zone 1 is the epitome of many interventions: it has been the focal point of Parisian policy interventions over the past decade. It was also one of the first zones to implement measures such as car-free Sundays in 2020, representing a ``progressive transformation'' of the city. 
Zone 2, located south of the Seine, has also undergone significant revitalization efforts, embodying transformation endeavors that reflect broader city improvements.
Zone 3 is a unique combination of districts from the center, east, and south of the city, and is an example of the results of a random combination of districts, found in the test set, to provide a diverse perspective.

\subsubsection{Visual Insights}
In Figures \ref{fig:zone_1_comparison}, \ref{fig:zone_2_comparison}, and \ref{fig:zone_3_comparison}, we compare the simulation output with the predictions of our model. The entire Paris street network is depicted in each image. The zone where a capacity reduction was implemented is outlined in black. The linewidth indicates the size of the street: trunk and primary streets have a thickness of 5, secondary streets have a thickness of 3, tertiary streets have a thickness of 2, and all other streets have a thickness of 1. The color indicates the percentage difference in car volume compared to the base case. For each zone, we used the same scale for both the actual simulation output and our predictions: -3\% to 3\% for Zones 1 and 2, and -5\% to 5\% for Zone 3. Values falling outside this range are represented by the color at the nearest scale boundary.

In all zones, the following observations can be made:
Overall, it is evident that our model can learn the changes in car volume. The trend (whether positive or negative, represented in the plot by red or blue) is recognizable on most streets.  The largest deviations between actual and predicted values are that the trends in the actual simulation output are usually stronger. This can be clearly seen as the colors in the ``actual predictions'' are more pronounced. It is also noticeable that the GNN finds it easier to identify a reduction in car volume compared to an increase in car volume. This can be explained by reductions in car volume occuring primarily on roads where the policy was directly implemented, making it a more straightforward pattern to detect, while increases in traffic volume yielding from spillovers and/or relocation of traffic by agents taking different routes.
Note also that the magnitude of change in car volume (-3 to 3\%, resp. -5 to 5\%) is not particularly high, probably posing challenges for the model. 

\subsubsection{Numerical Insights}

\begin{table}[htbp]
    \centering
    \caption{Numerical insights for Figure \ref{fig:zone_1_comparison} (Zone 1), \ref{fig:zone_2_comparison} (Zone 2) and \ref{fig:zone_3_comparison} (Zone 3)}
    \label{tab:result_table_graphics}
    \begin{tabularx}{\textwidth}{ p{1cm} | p{2cm} | p{2cm} | p{2cm} | X | X | X | X }
    \toprule
    \textbf{Zone} & \textbf{Length cap. reduction (km)} & \textbf{Network Scope} & \textbf{Road scope \protect\footnotemark} & \textbf{Variance} & \textbf{MSE: Baseline} & \textbf{MSE: Model} & \textbf{$R^2$: Model} \\
    \toprule
    \multirow{3}{*}{1} & \multirow{3}{*}{37} & \multirow{3}{*}{Paris} & All Roads & 6.45 & 0.54 & 1.01 & -0.87 \\
    \cmidrule(lr){4-8}
    & & & P & 28.66 & 1.64 & 2.34 & -0.42 \\
    \cmidrule(lr){4-8}
    & & & PST & 13.91 & 1.02 & 1.59 & -0.57 \\
    \cmidrule(lr){2-8}
    & \multirow{3}{*}{37} & \multirow{3}{*}{Zone 1} & All Roads & 50.61 & 2.48 & 1.62 & 0.34 \\
    \cmidrule(lr){4-8}
    & & & P & 84.12 & 6.11 & 6.05 & 0.01 \\
    \cmidrule(lr){4-8}
    & & & PST & 78.02 & 5.32 & 3.83 & 0.28 \\
    \midrule
    \multirow{3}{*}{2} & \multirow{3}{*}{83} & \multirow{3}{*}{Paris} & All Roads & 3.57 & 0.47 & 1.08 & -1.23 \\
    \cmidrule(lr){4-8}
    & & & P & 14.89 & 1.37 & 2.25 & -0.64 \\
    \cmidrule(lr){4-8}
    & & & PST & 7.00 & 0.86 & 1.67 & -0.92 \\
    \cmidrule(lr){2-8}
    & \multirow{3}{*}{83} & \multirow{3}{*}{Zone 2} & All Roads & 13.71 & 1.19 & 1.35 & -0.14 \\
    \cmidrule(lr){4-8}
    & & & P & 59.37 & 5.22 & 4.58 & 0.12 \\
    \cmidrule(lr){4-8}
    & & & PST & 25.40 & 2.38 & 2.64 & -0.11 \\
    \midrule
    \multirow{3}{*}{3} & \multirow{3}{*}{234} & \multirow{3}{*}{Paris} & All Roads & 358.20 & 4.10 & 2.25 & 0.45 \\
    \cmidrule(lr){4-8}
    & & & P & 623.00 & 9.91 & 5.11 & 0.48 \\
    \cmidrule(lr){4-8}
    & & & PST & 313.09 & 6.00 & 3.44 & 0.43 \\
    \cmidrule(lr){2-8}
    & \multirow{3}{*}{234} & \multirow{3}{*}{Zone 3} & All Roads & 455.51 & 6.40 & 3.00 & 0.53 \\
    \cmidrule(lr){4-8}
    & & & P & 1503.17 & 21.05 & 8.38 & 0.60 \\
    \cmidrule(lr){4-8}
    & & & PST & 785.49 & 12.13 & 5.80 & 0.55 \\
    \bottomrule
    \multicolumn{8}{l}{\small{``P'' stands for ``only Primary roads'', and ``PST'' stands for ``only Primary, Secondary, and Tertiary roads''}}\\
    \end{tabularx}
\end{table}

In Table \ref{tab:result_table_graphics}, we numerically evaluate the predictions for policies in the three example zones using the metrics MSE and $R^2$. As in Table \ref{tab:result_table}, we present the length of capacity reductions in kilometers, along with details on network and road scope, variance, baseline MSE, model MSE, and the model's \(R^2\). The network scope specifies whether the assessment covers all roads in Paris or is limited to those within the selected zone. The road scope indicates whether predictions are made for all roads, only primary roads, or exclusively primary, secondary, and tertiary roads.

In a first observation, the variance for the exemplary use cases for zone 1 and 2 is considerable lower compared to the variance in the whole data set (Table~\ref{tab:result_table}
A small baseline MSE, together with a low variance suggests that the policy's impact in these use cases is small in comparison to the whole data set likely due to a bias of rather large zones. Consequently, our model struggles to learn a clear policy signal. The resulting decreased accuracy in predictions, also observable in Figure~\ref{fig:zone_1_comparison} and \ref{fig:zone_2_comparison} provide a limitation of the current method. Only for primary roads with policy applied, a positive $R^2$ can be observed. 

Predictions for zone 3 on the other hand provide better results. The model's best performance is seen for primary roads, where it achieves a positive $R^2$ of 0.60, indicating a substantial improvement over the baseline. 
This underlines that the model is most effective in scenarios with higher variance, i.e. more substantial capacity changes, such as when capacity reductions are implemented across multiple zones.

\section{Conclusion} \label{sec:conclusion}
This paper presents a first approach to applying Graph Neural Networks (GNNs) for predicting changes in car volume due to policy interventions, serving as a ``Proof of Concept''. The results showed, it is generally possible to approximate the output of agent-based demand simulations using GNNs as in most scenarios significant improvement compared to a baseline can be observed. On the overall test set, our model achieves an MSE of 2.35 compared to a baseline of 4.0. For $R^2$, the baseline is 0, and our model achieves an accuracy of 0.41. Nevertheless, when policies are applied in exemplary smaller regions of the overall network, the model fails to predict policy effects accurately. On the positive side, the model performs generally better on higher road classes like primary and secondary road sections, which are often targeted, when policies are applied.

There are several challenges that are mandatory to be addressed in future work: First, due to constraints in computational time, we used down-sampled simulation representing only 0.1\% of all households for this study, resulting in inherent stochastic noise in the training data. Lower sub-sampling rates and/or higher sample sizes are expected to improve the results. Second, the generation of simulation scenarios used for training the GNN can likely be improved. Our training included relatively large combinations of neighboring Arrondissements (see Section \ref{sec:case_study}). The randomness of the selection process likely introduced a bias toward learning the effects of policies in larger district combinations. Third, in this study the whole network of Paris with over 31.000 edges is represented by the GNN. Further experiments have to be conducted whether only a subset of these sections is necessary to be represented to reduce the number of parameters in the model. Forth, the integration of further input features like spatial demand distributions or built environment might help learning the effects of policies.

Further, our machine learning model can undoubtedly benefit from further fine-tuning, which falls under the ``Model Development'' phase of machine learning. 
This includes, for example, more training data, additional loss terms and different evaluation metrics. For instance, one could incorporate different loss terms that can help capture the complexity of the task at hand, to see they change the learning process and whether they improve the model. Regarding evaluations, while MSE and $R^2$ are standard evaluation metrics for regression tasks, they may not be the best fit for our specific use case. Exploring alternative metrics could provide more insights about model's behaviour, which can lead to better proposed solutions. Finally, as discussed, increasing the amount of available data always improves model performance. While both computational and time expensive, increasing data richness - understood as percentage of population modelled - and data size, can give rise to better predictive performance.
Further, Graph Neural Networks are an active area of research in machine learning, and the available resources, such as implemented layers in PyTorch, are continually improving. Leveraging these advancements can further enhance the performance and capabilities of our model, making it more effective in predicting the impacts of various policies.

In the long run, the goal is to provide a generalizable model capable of aiding urban planners and policymakers in designing effective traffic management strategies, which could be achieved for example by coupling the model to an optimization framework selecting the best policies to achieve a certain objective function. The GNN based model would allow fast evaluations of policies, enabling testing a magnitude of different policy scenarios.
To achieve this goal, more policies beyond capacity reduction, e.g. road tolls or improvements in public transport, have to be included in the training data. Additionally, effects on other modes of transport have to be incorporated, for example by adding a public transport layer to the model.

\section{Acknowledgments}
We thank the German Federal Ministry of Transport and Digital Infrastructure for providing funding through the project ``MINGA'' with grant number 45AOV1001K. We remain responsible for all findings and opinions presented in the paper.

We thank Alejandro Tejada Lapuerta from Helmholtz Munich for fruitful discussions about setting up the model and interpreting the results. Further, we thank Dominik Fuchsgruber from the TUM School of Computation, Information and Technology for helpful discussions on finetuning the Graph Neural Network. 

Chat-GBT 4.0 was used for summarizing paragraphs. 

\section{Author contributions}

The authors confirm their contributions to the paper as follows: E. Natterer, R. Engelhardt, S. Hörl, and K. Bogenberger conceived and designed the study. S. Hörl provided the MATSim simulation for Paris. E. Natterer executed the MATSim simulations and developed the model. E. Natterer and R. Engelhardt analyzed and interpreted the results. All authors reviewed the results and approved the final version of the manuscript.



\bibliographystyle{trb}
\bibliography{references,sebastians_bib}

\end{document}